\documentclass[runningheads]{llncs}
\usepackage[T1]{fontenc}
\usepackage{url}
\usepackage{cite}
\usepackage{ifsym}
\usepackage{float}
\usepackage{xcolor}
\usepackage{bbding}
\usepackage{caption}
\usepackage{colortbl}
\usepackage{booktabs}
\usepackage{hyperref}
\usepackage{graphicx}
\usepackage{adjustbox}
\usepackage{algorithm}
\usepackage{textcomp}
\usepackage{algorithmic}
\usepackage{amsmath,amssymb,amsfonts}

\newcommand{\myorcidID}[1]{\href{https://orcid.org/#1}{\includegraphics[width=8pt]{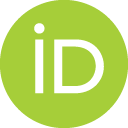}}}
\definecolor{customblue}{RGB}{135, 179, 224}
\begin{document}
\title{Plug-and-Play Performance Estimation for LLM Services without Relying on Labeled Data}
%
\titlerunning{Plug-and-play Performance Estimation for LLM Services}
%
\author{Can Wang\myorcidID{0009-0000-4139-6855} \and
Dianbo Sui \and
Hongliang Sun \and
Hao Ding \and
Bolin Zhang \inst{(}\Envelope\inst{)} \and
Zhiying Tu \inst{(}\Envelope\inst{)}}
\authorrunning{C. Wang et al.}

%
\institute{Harbin Institute of Technology, Harbin, Heilongjiang, China\\
\email{\{23B903072\}@stu.hit.edu.cn}
\email{\{suidianbo\}@hit.edu.cn}
\email{\{21B903094, dinghao\}@stu.hit.edu.cn}
\email{\{brolin, tzy\_hit\}@hit.edu.cn}}

\maketitle              

\begin{abstract}
Large Language Model (LLM) services exhibit impressive capability on unlearned tasks leveraging only a few examples by in-context learning (ICL).
However, the success of ICL varies depending on the task and context, leading to heterogeneous service quality. Directly estimating the performance of LLM services at each invocation can be laborious, especially requiring abundant labeled data or internal information within the LLM. 
This paper introduces a novel method to estimate the performance of LLM services across different tasks and contexts, which can be "plug-and-play" utilizing only a few unlabeled samples like ICL.
Our findings suggest that the negative log-likelihood and perplexity derived from LLM service invocation can function as effective and significant features. Based on these features, we utilize four distinct meta-models to estimate the performance of LLM services. Our proposed method is compared against unlabeled estimation baselines across multiple LLM services and tasks. And it is experimentally applied to two scenarios, demonstrating its effectiveness in the selection and further optimization of LLM services.

\end{abstract}
%
%
%
\section{Introduction}

\label{sec:intro}
Large language models (LLM) have the capability to understand and generate natural language text, making them valuable tools for a variety of natural language processing tasks such as text generation, translation~\cite{trans}, summarization~\cite{summar}, question answering~\cite{qa}, and more. LLM services, such as OpenAI LLM API~\footnote{OpenAI publishes LLM services through \url{https://openai.com/blog/openai-api}}, allow users to conveniently solve their tasks by interacting with LLM in a flexible conversational manner, without needing to know whether the LLM has been trained on these tasks or not.

This remarkable capability is realized through the paradigm of In-Context Learning (ICL)~\cite{BrownMRSKDNSSAA20}, which enables the LLM to generalize
rapidly only employing a few labeled examples without requiring additional training. 
However, such a paradigm is not flawless. Many studies have revealed such a reality: ICL is highly sensitive to task and context~\cite{zhao2021calibrate, perez2021true}. 
ICL can demonstrate significant advantages in certain tasks when using appropriate LLM services and contexts, such as solving entity linking tasks with the Phi-2 service. But it can be virtually ineffective in other scenarios, like solving web-question tasks with the Llama-7B service\footnote{
We conducted experiments to prove this. For more detailed information about this paper, including the dataset, hyperparameter settings, etc., please see: \url{https://github.com/WangCan1178/Plug-and-Play-Estimation}}. 
Therefore, in the face of different tasks and contexts, it is both challenging and necessary to estimate the performance of LLM services in advance.


To estimate the LLM services' performance, typical solutions use labeled data to invoke LLM services~\cite{abs230506474, singhal2022assessing}, which necessitate collecting labels and testing them for each task. 
However, generally purposed LLM services address a wide range of natural language tasks and most of these tasks are not human-labeled. 
Especially in domain-specific tasks that require expertise, such as medical or law text understanding, the high cost of annotation poses a significant challenge.
Another solutions avoiding labeled data by exploiting the information within the LLM during inference, potentially requiring the LLM’s architecture and parameters to be open~\cite{HuangWLWSWYZ23, li2022estimating,GargBLNS22,10207589}.
These approaches have limitations for LLM services that only provide usage access without disclosing internal information, and in practice, extracting internal information on a large number of heterogeneous LLM services is also a time-consuming and laborious work.

Building on this, we explore a more practical and appealing idea to estimate the performance of LLM services, which can be "plug-and-play" for various LLM services and unlabeled tasks in different contexts.
In detail, we explore the common relationship between the semantic features exhibited during the invocation of LLM services and performance. 
We find two useful features, negative log-likelihood and perplexity, which rely solely on the answers generated during LLM service invocation, but can reflect the performance potential of the LLM service on the current task and context.
Then, we propose our meta-model based approach tightly integrated with the ICL paradigm: for a quick and reliable performance estimation, only the answers of the LLM service to a few examples are needed. Our novel training and inference approach using linear interpolation makes the meta-model effective and generalizable, which can be used for a wide range of different LLM services and unlabeled tasks without retraining.

The contributions of this paper are:

\begin{itemize}
\item We explore the common phenomenon exhibited during the invocation of LLM services, and select the available features based on their relevance.
\item We propose a method for LLM service performance estimation that is able to reach low-error estimates at little cost on various unlabeled tasks, which can be applied to most LLM services that do not know the internal information.
\item We verify the effectiveness of our method in two scenarios: the selection of LLM services and the further optimization for few-shot tasks of LLM services, proving that it can be helpful in various future works.
\end{itemize}


\begin{figure*}[htbp]
\centerline{\includegraphics[width=0.87\linewidth]{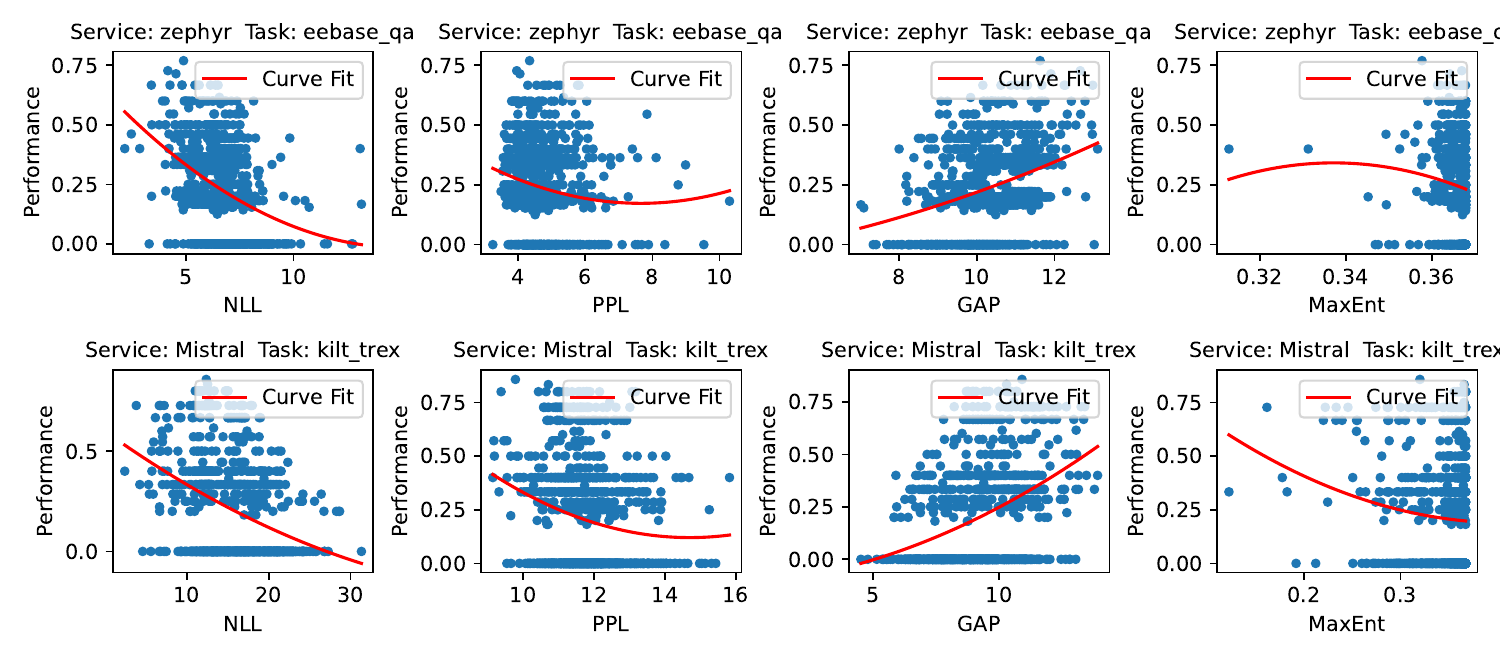}}
\caption{Distribution of the four features and the LLM service performance, as well as the fitting curve (from two randomly selected task invocation results).}
\label{fig:case}
\end{figure*}

\section{Pilot Experiments}


\label{sec:pil}
This section describes the phenomenon we observe when LLM services are invoked to perform ICL. 
Specifically, we mainly explore the following two research questions:  \textbf{RQ1}: What features can be extracted on unlabeled task when invoking LLM services through ICL paradigm? \textbf{RQ2}: How to select appropriate features to reflect the performance of LLM services?
%

\subsection{Experiment Setup}
In the pilot experiments, we select the top 5 generative LLM services according to the downloads of hugging face model library (\url{https://huggingface.co/models}), which have different sizes and structures. Besides, 
we choose a representative benchmark dataset: \textsc{CrossFit}~\cite{YeLR21}, a benchmark to study the generalization capability of LLMs, containing 160 different few-shot NLP tasks.
Given that our method is based on ICL and few-shot tasks, we don't use datasets for common NLP tasks that LLM services may have seen the data during training.
We sample examples from the training dataset of each task, constitute the context of the unlabeled data, and invoke the LLM service on the testing dataset.
And F1-score~\cite{f1score} is used to calculate the accuracy of the generation on unlearned tasks, reflecting the performance of LLM service when invoked.

\subsection{What Features Can be Extracted on Unlabeled Task? (RQ1)}

\label{subsec:fea}
To answer the RQ1, we survey LLM services on the market and find that almost all of them provide (top-few) word-list probabilities of the reasoned answer. We use this probability to come up with the following usable features. 
A intuitive illustration of their strong correlation can be observed in Figure~\ref{fig:case}, by fitting the distribution between these features and performance
on these tasks.


\textbf{Negative log-likelihood (NLL).} 
We treat the process of the LLM service invocation as a generation task. NLL can be used for measuring how well LLM fit on the dataset, which can be obtained from each generated sequence as following: 
\begin{equation}
nll(x) = -\sum_{t=1}^{|x|} \log P(x_{t} \mid x_{<t}; \theta)\
\end{equation}
where $x$ is the output sequence of the LLM service with parameter $\theta$, and $P(x_{t} \mid x_{<t})$ is the maximum probability assigned at $t$-th token. The smaller the value, the more confident about the generated sentence.

\begin{figure}[!t]
\centerline{\includegraphics[width=0.35\linewidth]{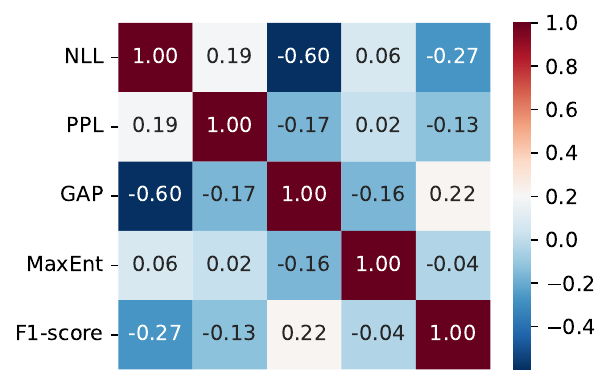}}
\caption{Pearson correlation coefficient of features and LLM services performance.}
\label{fig:heatmap}
\end{figure}

\textbf{Perplexity (PPL).} 
Perplexity reflects the likelihood of a LLM having seen and learned (in other word, be pretrained on) this data before. It is calculated based on the probability that the LLM reconstructs the input sequence.
\begin{equation}
ppl(x) = \exp{(-\sum_{t=1}^{|x|} \log P(\Tilde{x_{t}} \mid x_{<t}; \theta))}\
\end{equation}
Its calculation is similar to NLL, except the predicated token $x_{t}$ is replaced by the input token $\Tilde{x_{t}}$ in the conditional probability. The smaller the value, the more likely it is that the LLM has seen and learned the generated sentence.

\textbf{GAP.} 
GAP is defined as the difference between the probability of the most likely token (i.e., the first ranked token) and the probability of the second most likely token (i.e., the second ranked token) in the probability distribution generated for the current word. 
\begin{equation}
gap(x) = \sum_{t=1}^{|x|} P(x_{t} - x_{t\_sec} \mid x_{<t}; \theta)\
\end{equation}
GAP takes into account the effect of potentially possible answers. And the bigger the value, the more accurate about the generated sentence.

\textbf{Maximum Entropy (MaxEnt).} 
Entropy is an indicator to measure the uncertainty about the generated tokens. Preliminary experiments~\cite{fadeeva2024factchecking} show that simply taking the maximum token entropy significantly outperforms other aggregation methods such as averaging or taking the minimum. 
\begin{equation}
MaxEnt(x) = \max_{t \in |x|} \mathcal{H}(x_{t} \mid x_{<t}; \theta)\
\end{equation}
where $ \mathcal{H}(x_{t} \mid x_{<t}; \theta)$ is the entropy of the current token calculated from its probability. The smaller the value, the more certain about the generated sentence.

\subsection{How to Select Appropriate Features to Reflect the Performance of LLM Services? (RQ2)}

Based on the proposed features in Section~\ref{subsec:fea}, we can compute the correlation between these features and the LLM services performance, in units of tasks.
Pearson correlation coefficient is used as the indicator of measurement. 
And F1-score is used to calculate the accuracy of the generation on unlearned tasks, reflecting the performance of LLM service. 
Figure~\ref{fig:heatmap} shows the overall correlations by simply taking the average over all results.
These features can constitute different combinations that reflect the LLM service performance. According to the theory of correlation and collinearity~\cite{corring,hall1999correlation}, we expect to select combinations that are strongly correlated with the performance, but not strongly correlated with other features. Thus, we define a score to this end as followed.

\begin{equation}
Score(F) = \sum_{f_i, f_j\in F} corr(f_i,F1) - corr(f_i,f_j)
\end{equation}

where $F=\{f_1, f_2, ... f_n\}$ is a combination of different features, and $corr(f_i,f_j)$ denotes the correlation between the feature $f_i$ and $f_j$, without repeating calculating $corr(f_j, f_i)$. Through this score, we find the best combination of features is $F=\{NLL, PPL\}$, denoting these two features can reflect the performance of LLM service best from two different aspects. 


\begin{figure*}[t]
\centerline{\includegraphics[width=0.95\linewidth]{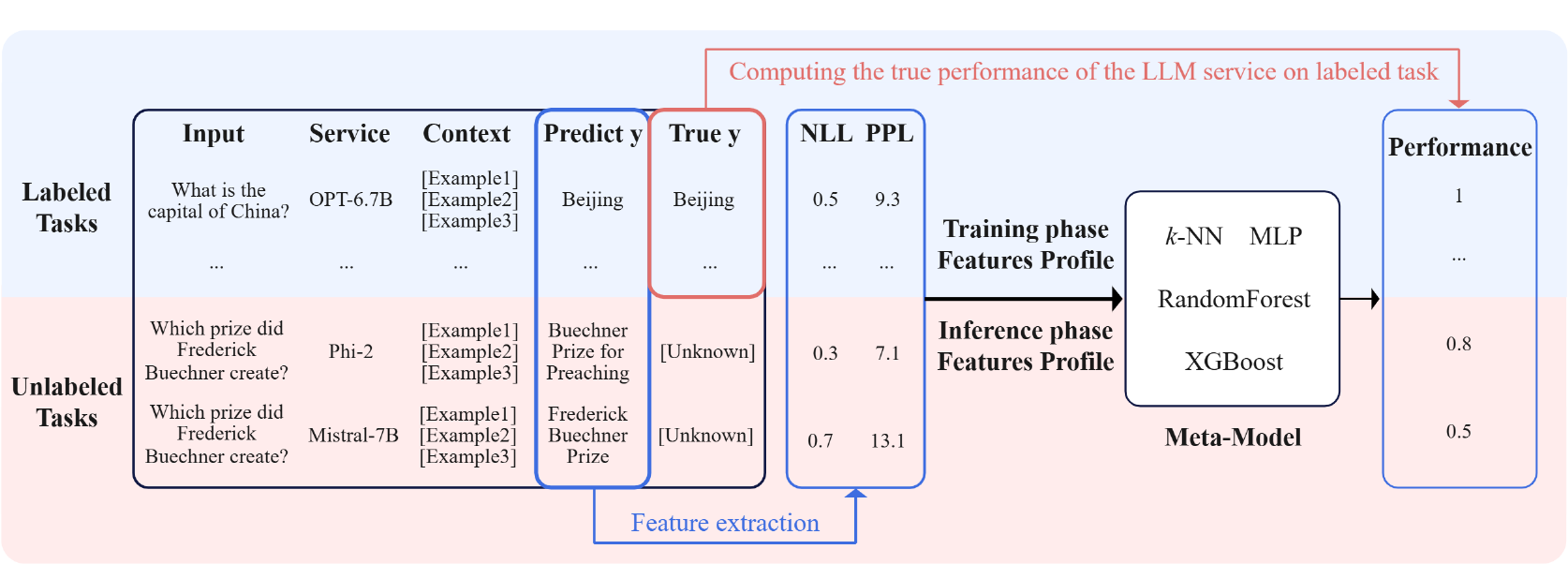}}
\caption{Procedure of our meta-model based LLM service performance estimation.}
\label{fig:meta}
\end{figure*}

\section{Methodology: LLM Services Performance Estimation}

\label{sec:llm}
In this section, we first illustrate the definition of LLM services performance estimation problem in subsection~\ref{subsec:pro}. 
Then, the meta-model based method we proposed is introduced in subsection~\ref{subsec:train}, detailing its novel training and inference process in subsection~\ref{subsec:inf}. The whole procedure of the proposed meta-model based method is shown as shown in Figure~\ref{fig:meta}.

\subsection{Problem Definition}

\label{subsec:pro}

In this paper, we focus on estimating performance of LLM services on unlearned and unlabeled tasks. Our goal can be formalized as investigating how to quickly and cheaply estimate the performance $\widehat{per}^{S}_{T,C}$ of ICL, given a LLM service $S$, an unlearned and unlabeled task $T = \{x^{(1)}, \dotsc, x^{(n)}\}$, with the context $C$. 

Absolute error is used to measure the effectiveness of our method, defined as $|per^{S}_{T,C} - \widehat{per}^{S}_{T,C}|$, where $per^{S}_{T,C}$ is the true performance of the LLM service $S$ invoked to handle the same task $T$ with $C$. 
We explore performance estimation on a broad LLM service market $\{S_i\}_{i=1}^{I}$, a wide range of natural language tasks $\{T_j\}_{j=1}^{J}$ and different contexts $\{C_k\}_{k=1}^{K}$. Therefore, the final mean abstract error (MAE) is calculated by the following method, using the average absolute error:

\begin{equation}
MAE = \frac{1}{I}\frac{1}{J}\frac{1}{K} \sum_{i \in I}\sum_{j \in J}\sum_{k \in K}|per^{S_i}_{T_j,C_k} - \widehat{per}^{S_i}_{T_j,C_k}|\
\label{equ:err}
\end{equation}

In different subsequent works, several of the terms in the Equation~\ref{equ:err} can be fixed to simplify the calculation. Such as selecting the most appropriate LLM service for a specific fixed task with a given context, it only needs to minimize the average error of all LLM services.

\subsection{Meta-Model Based Method}
\label{subsec:train}

Meta-model~\footnote{Extend from metamodeling(~\url{https://wikipedia.org/wiki/Metamodeling})} refers to a high-level model that does not predict the data directly, but makes the final prediction or decision by combining and analyzing the prediction results of other base models. In our proposed method, the meta-model is able to accept the features from generated answers when invoking the LLM service, and to estimate the performance when inference.

For efficiency, we choose meta-models with simple structures containing much fewer parameters than LLM, which are easy to train and have fast inference speed. Four meta-models with different architectures are selected: \textbf{$k$-Nearest Neighbors} ($k$-NN), which estimates the performance of the LLM service by measuring the similarity of features between samples~\cite{knn}. \textbf{Multilayer Perceptron} (MLP), which estimates the performance of LLM service by the probability propagation between neurons~\cite{mlp}. \textbf{RandomForest}, which estimates the performance of LLM service by bagging multiple weak Learners~\cite{for}. 
\textbf{eXtreme Gradient Boosting} (XBoost), which estimates the performance of LLM service by boosting multiple weak Learners~\cite{xgb}. 

\subsection{Training and Inference}

\label{subsec:inf}
The goal of the training phase is to obtain a good meta-model $M$, which must possess sufficient generalization capability to be applied across a wide range of LLM services and open-domain tasks. It is satisfied through three required inputs: a set of LLM services $\{S_i\}_{i=1}^{I}$, a set of labeled tasks $\{T_j^{labeled}\}_{j=1}^{J}$, and a set of contexts $\{C_k\}_{k=1}^{K}$ sampled from the respective task. 
These three inputs can be arbitrarily combined to obtain the results of invocations of different LLM services under various tasks and contexts. 
As mentioned in Section~\ref{sec:pil}, we can extract the useful meta-model features $nll^{S_i}_{T_j^{labeled},C_k}$ and $ppl^{S_i}_{T_j^{labeled},C_k}$ of the series of invocations, as well as the true performance $per^{S_i}_{T_j^{labeled},C_k}$ demonstrated.


A key problem is that different scale sizes of tasks result in different dimensions of features. We borrow the idea of profile~\cite{FuYXRJ23}, and map the features to the same dimension $d$ by linear interpolation. It is defined as follows, where $D$ is the dataset size of the current task $T_j^{labeled}$, and $f_n$ is the $n$-th index of the features such as $nll^{S_i}_{T_j^{labeled},C_k}$ or $ppl^{S_i}_{T_j^{labeled},C_k}$.

\begin{equation}
f_n = liner(f_{\lfloor|D|\times n/d\rfloor}, f_{\lceil|D|\times n/d\rceil})\
\end{equation}

In this way, by continuously reducing the difference between the estimated performance $\widehat{per}^{S_i}_{T_j^{labeled},C_k}$ and the true performance ${per}^{S_i}_{T_j^{labeled},C_k}$, the meta-model gradually converges to the point where it can estimate the LLM service performance on different unlearned tasks. 


In the inference phase, a partial subset of the unlabeled task $T^{*unlabeled} \subseteq T^{unlabeled}$ is selected, and the estimated LLM service $S$ is invoked with the context $C$ for inference. 
Using the same method as in the training phase, the features $nll^{S}_{T^{*unlabeled},C}$ and $ppl^{S}_{T^{*unlabeled},C}$ and are obtained and mapped to a $d$ dimensional space. Then, the trained meta-model $M$ is applied and the estimated performance of LLM service is obtained as defined.
 
\begin{equation}
\widehat{per}^{S}_{T^{*unlabeled},C} = M(nll^{S}_{T^{*unlabeled},C}, ppl^{S}_{T^{*unlabeled},C})\
\end{equation}

In summary, our trained meta-model achieves estimation on a wide range of unlabeled tasks and contexts, by exploring the relationship between the features exhibited during LLM service invocation and its performance.

\section{Experiments}

\label{sec:exp}
In this Section, we first introduce the baseline methods, including methods using both labeled and unlabeled data. Then in subsection~\ref{subsec:exp}, the details of our experiments are presented. The main results are given in subsection~\ref{subsec:res}, demonstrating the effectiveness and practicality of our approach. Finally, we conduct an ablation study to show the impact of features and the number of unlabeled samples used to make estimations.

\subsection{Baselines}

We design the following baselines that do not involve LLM internal information to compare with our method, which allows LLMs to be used as black-box services.

\textbf{Sample accuracy of labeled examples} (\textsc{Sample$^n$}). 
It is straightforward to estimate the performance of LLM services exhibited in different tasks by labeling the few data~\cite{singhal2022assessing}. 
This method samples $n$ examples from the dataset of the task $T^{unlabeled}$ to label, and calculate the accuracy of these $n$ examples as the performance of the whole task, which we call it \textsc{Sample$^n$}. 
According to the law of large numbers, the more sample examples are labeled, the closer the estimated accuracy is to the true performance, at the cost of more expensive labeling costs.
When estimating LLM service performance, we want our method to be able to approximate the accuracy of Sample$^n$ without using labeled data.

\[\widehat{per}^{S}_{T,C} (\textsc{Sample$^n$}) = \frac{1}{n}\frac{1}{K} \sum_{i=1}^{n} \sum_{k \in K} per^S_{x^{(i)}, C_k}\]

where $per^S_{x^{(i)}, C}$ is the true performance of the LLM service $S$ invoked on the labeled sample $x^{(i)}$ sampled from the unlabeled task $T$ dataset.

\textbf{Average accuracy on the training dataset} (\textsc{AvgTrain}). 
Similarly, the average over all labeled tasks can also be used as a baseline for LLM service. 

\[\widehat{per}^{S}_{T,C} (\textsc{AvgTrain}) = \frac{1}{J}\frac{1}{K} \sum_{j \in J}\sum_{k \in K}per^{S}_{T_j,C_k}\]

\textbf{Average threshold of confidence} (\textsc{ATC}). 
Another practical baseline approach is to obtain a threshold based on the confidence of LLM service on task-level~\cite{GargBLNS22}, whereby the accuracy is predicted by the proportion of unlabeled instances where model confidence surpasses the threshold. And the average of the \textsc{ATC} obtained from the estimates of each seen task is used as a baseline.
\[\widehat{per}^{S}_{T,C} (\textsc{ATC}) = \frac{1}{J}\frac{1}{K} \sum_{j \in J}\sum_{k \in K}{atc}^{S}_{T_j,C_k}\]

where $atc^{S}_{T_j,C_k}$ is the accuracy estimation of task $T_j$ using the threshold of model confidence.

\subsection{Experimental Details.}

\label{subsec:exp}
We use the dataset mentioned in Section~\ref{sec:pil} because it provides enough few-shot NLP tasks to facilitate the study of performance of LLM services across tasks. Thirteen open-ended generation tasks are selected for our experiments.
We combine the "train" and "dev" set to train our meta-model as labeled tasks. For each task, three examples are sampled from it at a time as context. And we conducted a total of 50 different sampling times to fully investigate the impact of context on the invocation of different LLM services. In total, we have experimented with executing 5 LLM services on 13 tasks with 50 contexts (3250 ICL settings).

We use 1000 unlabeled samples and 5-fold cross-validation to verify our method's effectiveness. The number of features $d$ is set to 100, and the optimal hyperparameters are selected by grid search for each meta-model.

\subsection{Main Results.}
\label{subsec:res}
In this subsection, we first compare the error of our proposed method and baselines. Then we demonstrate the effectiveness of our method in terms of the accuracy improvement of two subsequent works.

\textbf{LLM services performance estimation performs all unlabeled baselines.}
Table \ref{tab:std} shows the MAE obtained by performance estimation on different LLM services.
On all LLM services, our method performs better than the unlabeled baselines on average across the 13 tasks. And in the best case (XGBoost with OPT-6.7B), the meta-model's MAE is 31.1\% lower than the best baseline method without labels.

Compared with method \textsc{Sample$^n$} using labeled data, our method can mostly outperform the method with 16 samples, and sometimes even outperform the method with 32 samples.
In particular, when not differentiated by LLM service, our model performs at a comparable level to sampling 64 samples(4.78 $\pm$ 2.10), which shows the high generalization of our method. The implication is that we can significantly save annotation costs to estimate the performance of different LLM services on a wide range of natural language tasks.
Furthermore, note that the standard deviation of the best meta-model on the LLM performance estimation using our method is significantly smaller than that of the other method. It indicates the stability of our method, that is, it is almost unaffected by the sampled different unlabeled samples.

\begin{figure*}[t]
\centerline{\includegraphics[width=0.82\linewidth]{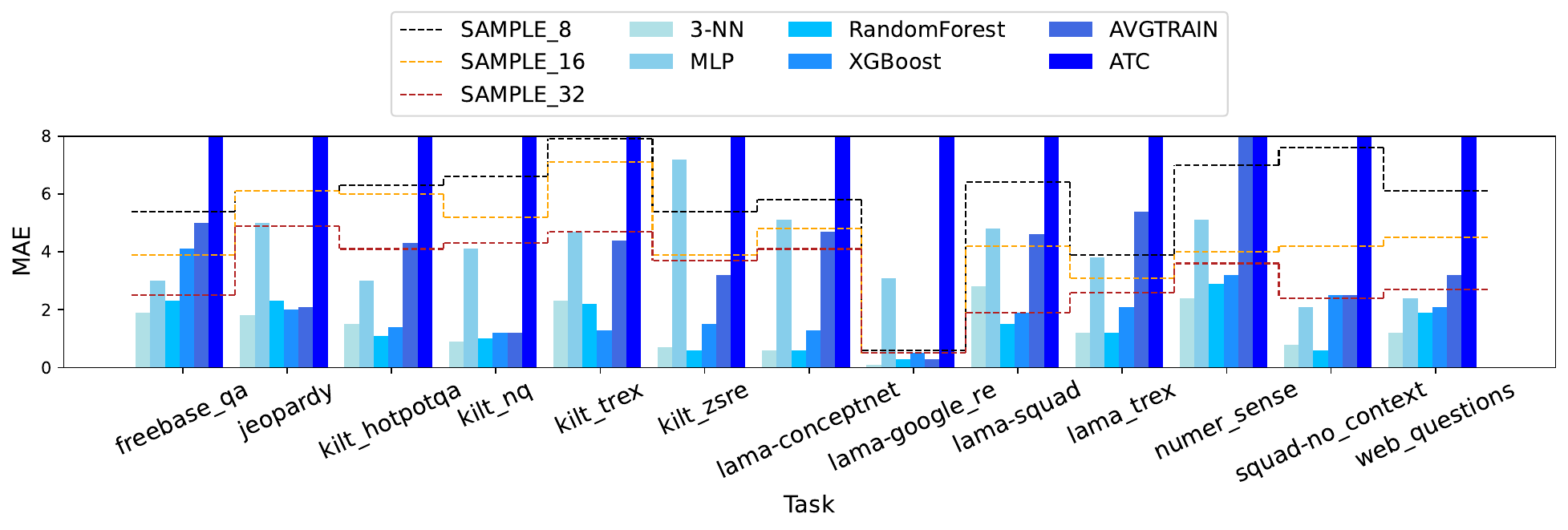}}
\caption{Experimental results (MAE) for different tasks of the LLM services performance estimation (our method) and baselines.}
\label{fig:result}
\end{figure*}

We also explore the performance of LLM service performance estimation on different tasks, and the results are presented in Figure~\ref{fig:result}. It shows that in all 13 few-shot tasks, our best meta-model outperforms all baseline methods, including the previous best performing method \textsc{SAMPLE}$^{32}$. This may be due to the fact that the probability distribution of the results for the same task is similar, which leads to faster convergence and better performance of our method.
It is more practical and attractive than estimating the LLM service performance on multiple tasks, because the performance estimation on a certain given task is more in line with the actual demands.

\begin{table}[!t]
\caption{Experimental results (MAE) and variations (SD) for different LLM services performance estimation on our method and baselines.}
\begin{center}
\begin{adjustbox}{width=0.80\textwidth}
\begin{tabular}{ccccccc}
\toprule
\textbf{LLM service} & \textbf{Llama-7B} & \textbf{Mistral-7B} & \textbf{OPT-6.7B} & \textbf{Phi-2} & \textbf{Zephyr-7B-$\mathrm{\beta}$} & \textbf{Total}\\
\midrule \midrule
\textbf{Baselines} & & & & & & \\
\textsc{Sample$^8$} & 7.98 $\pm$ 4.10 & 5.27 $\pm$ 1.34 & 10.23 $\pm$ 3.09 & 10.20 $\pm$ 3.79 & 4.74 $\pm$ 2.45 & 8.28 $\pm$ 4.25\\

\textsc{Sample$^{16}$} & 6.78 $\pm$ 2.82 & 4.12 $\pm$ 1.59 & 8.70 $\pm$ 2.75 & 9.22 $\pm$ 2.61 & 3.98 $\pm$ 1.49 & 6.12 $\pm$ 3.60\\

\textsc{Sample$^{32}$} & 4.15 $\pm$ 1.54 & 3.12 $\pm$ 0.94 & 6.24 $\pm$ 2.38 & 6.78 $\pm$ 2.41 & 3.10 $\pm$ 1.36 & 5.14 $\pm$ 2.84\\

\textsc{AvgTrain} & 6.20 $\pm$ 2.70 & 5.74 $\pm$ 4.64 & 8.96 $\pm$ 3.71 & 8.60 $\pm$ 2.23 & 5.80 $\pm$ 2.48 & 6.74 $\pm$ 4.62\\

\textsc{ATC} & 40.91 $\pm$ 10.24 & 38.82 $\pm$ 5.36 & 30.56 $\pm$ 11.42 & 31.10 $\pm$ 9.98 & 39.20 $\pm$ 9.02 & 39.82 $\pm$ 5.37 \\
\midrule
\textbf{Meta Models} & & & & & & \\

3-NN & 6.50 $\pm$ 2.50 & 7.30 $\pm$ 2.69 & 7.02 $\pm$ 2.63 & 7.42 $\pm$ 3.61 & 6.18 $\pm$ 0.23 & 7.30 $\pm$ 2.96 \\

MLP & 7.24 $\pm$ 5.60 & 5.30 $\pm$ 0.71 & 7.24 $\pm$ 0.82 & 6.58 $\pm$ 1.31 & 5.50 $\pm$ 1.44 & 5.30 $\pm$ 1.70 \\

RandomForest & 5.80 $\pm$ 0.99 & \textbf{4.02 $\pm$ 1.61} & 8.60 $\pm$ 1.60 & \textbf{5.38 $\pm$ 1.46} & \textbf{4.04 $\pm$ 1.34} & \textbf{4.72 $\pm$ 1.61} \\

XGBoost & \textbf{4.76 $\pm$ 0.63} & 5.42 $\pm$ 1.59 & \textbf{6.17 $\pm$ 0.81} & 5.44 $\pm$ 1.60 & 4.56 $\pm$ 1.17 & 5.42 $\pm$ 2.59 \\

\bottomrule
\end{tabular}
\end{adjustbox}
\label{tab:std}
\end{center}
\end{table}

\begin{figure}[t]
\centerline{\includegraphics[width=0.65\linewidth]{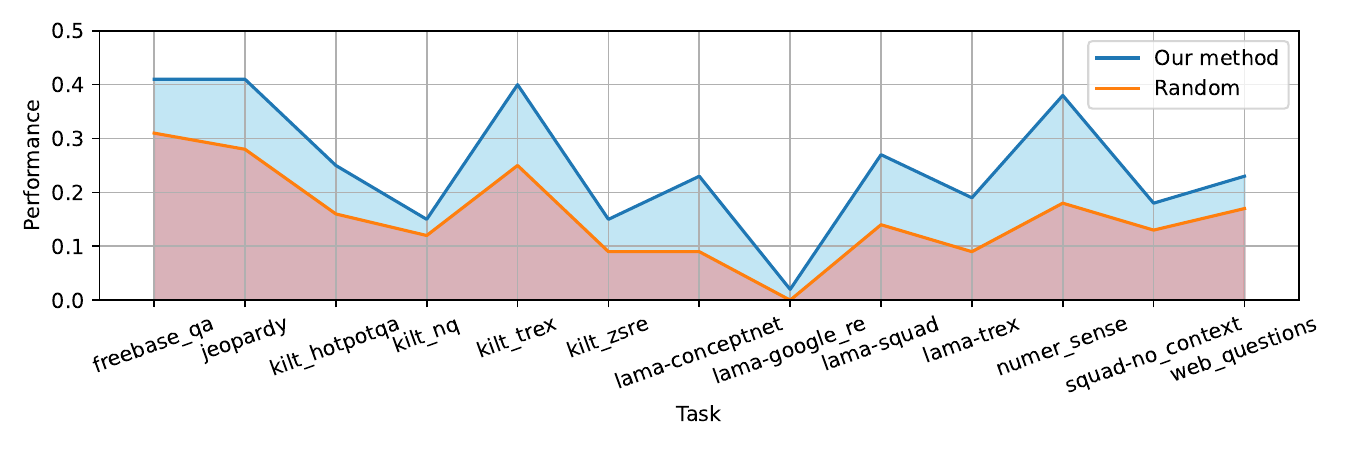}}
\caption{Execution performance under the settings of our method and randomly selected services or contexts.}
\label{fig:select}
\end{figure}

\textbf{LLM services performance estimation helps the subsequent works.} 
To verify the practicality of our method, we conduct experiments of two common scenarios, applying LLM services performance estimation to subsequent works.

The first scenario considers the selection of services and contexts when the user invokes the LLM service to perform some unlearned task.
We experiment with the best hyperparameters of the best meta-model architecture (RandomForest). Five LLM services and ten different contexts are randomly sampled for a total of 50 ICL settings. 
The estimation performance on different settings are given through the inference of the meta-model, which the best one are selected as the recommended service and context of our method.
It is compared with the random selection of services and contexts that often happens in the actual invocation scenario, and the results shown in Figure~\ref{fig:select} are obtained.

In 13 different few-shot tasks, the selected services and contexts using our proposed LLM service performance estimation indeed exhibit stronger ICL capability. This is undoubtedly appealing, as users can improve the performance of the current task by even up to 21\% at no additional annotation cost.

Another scenario considers domain tasks that are difficult for all LLM services, such as lama-conceptnet (concept question with answer is a single word), which performs best with only an F1-score of 0.12. At this time, further optimization of the LLM service is necessary, and a common approach is to fine-tune the LLM's parameters, which is often laborious and resource-consuming. Our approach can help indicate which LLM service has a wider optimization space, and to make a better choice in advance.

The reason we believe that estimated performance can represent the optimization space is that the LLM services are always under-fitting and low-performaning on these tasks.
The higher the estimated performance of the LLM service, the stronger its potential language modeling ability for the task, indicating it can perform best after further optimization.
Similarly, we performed the experiments in different 50 settings, and use the difference of performance to indicate how much the LLM service has improved after fine-tuning, which is defined as $diff = {per}^{S_{fineturn}}_{T,C} - {per}^{S}_{T,C}$. 


Table~\ref{tab:opt} presents the results (in the form of ${per}^{S}_{T,C} \pm diff$ to display the changes before and after fine-tuning) of further optimization on the three worst performing tasks. It is shown that the estimation of LLM service performance can subconsciously indicate the best suitable service for fine-turning. And it can provide useful guidance for further optimization of LLM services on low-performance tasks.

\subsection{Ablation Study}

\begin{table}[t]
\caption{Fine-tuning effects on the low-performing tasks of LLM services selected by our method and all LLM services.}
\begin{center}
\begin{adjustbox}{width=0.48\textwidth}
\begin{tabular}{cccccc} 
\toprule
 \textbf{Task} & \textbf{kilt\_zsre} & \textbf{lama-conceptnet} & \textbf{lama-google\_re} \\
\midrule \midrule
\textbf{Llama-7B} & 0.09 - 0.02 & \textbf{0.06 + 0.09} & 0.00 + 0.08 \\
\textbf{Mistral-7B} & \cellcolor{gray!25}\textbf{0.12 + 0.13} & 0.12 + 0.05 & 0.01 + 0.10 \\
\textbf{OPT-6.7B} & 0.04 + 0.05 & 0.09 + 0.02 & 0.01 + 0.02 \\
\textbf{Phi-2} & \textbf 0.03 + 0.06 & \cellcolor{gray!25}\textbf{0.22 + 0.09} & 0.00 - 0.01 \\
\textbf{Zephyr-7B-$\mathrm{\beta}$} & 0.12 + 0.07 & 0.05 + 0.03 & \cellcolor{gray!25}\textbf{0.02 + 0.11} \\
\bottomrule
\end{tabular}
\end{adjustbox}
\label{tab:opt}
\end{center}
\end{table}

\begin{table}[!t]
\caption{Evaluation results (MAE) and variants (SD) of different features selected to use in our method.}
\begin{center}
\begin{adjustbox}{width=0.70\textwidth}
\begin{tabular}{ccccccc}
\toprule
\textbf{LLM service} & \textbf{Llama-7B} & \textbf{Mistral-7B} & \textbf{OPT-6.7B} & \textbf{Phi-2} & \textbf{Zephyr-7B-$\mathrm{\beta}$} & \textbf{Total}\\
\midrule
\midrule
\textbf{NLL only} & & & & & & \\

3-NN & 8.22 $\pm$ 3.50 & 8.15 $\pm$ 3.29 & 7.02 $\pm$ 2.63 & 10.67 $\pm$ 4.64 & 8.82 $\pm$ 2.23 & 9.10 $\pm$ 4.37 \\

MLP & 9.23 $\pm$ 5.51 & 7.03 $\pm$ 4.25 & 8.51 $\pm$ 2.46 & 7.70 $\pm$ 3.27 & 7.47 $\pm$ 3.40 & 6.52 $\pm$ 3.86 \\

RandomForest & 7.85 $\pm$ 2.24 & \textbf{6.11 $\pm$ 3.57} & 9.00 $\pm$ 3.97 & 7.42 $\pm$ 2.34 & \textbf{6.83 $\pm$ 3.57} & \textbf{6.90 $\pm$ 3.13} \\

XGBoost & \textbf{6.61 $\pm$ 1.90} & 7.94 $\pm$ 2.55 & \textbf{6.17 $\pm$ 3.24} & \textbf{7.32 $\pm$ 2.34} & 6.97 $\pm$ 3.76 & 7.17 $\pm$ 4.05 \\

\midrule

\textbf{PPL only} & & & & & & \\

3-NN & 14.50 $\pm$ 4.78 & 9.23 $\pm$ 4.31 & 12.78 $\pm$ 3.79 & 11.45 $\pm$ 5.82 & 9.14 $\pm$ 2.56 & 11.68 $\pm$ 4.56 \\

MLP & 12.67 $\pm$ 6.34 & 11.95 $\pm$ 5.13 & 9.72 $\pm$ 2.89 & 14.79 $\pm$ 5.14 & 10.56 $\pm$ 4.60 & 9.19 $\pm$ 4.02 \\

RandomForest & 10.97 $\pm$ 5.67 & 9.85 $\pm$ 4.98 & 11.14 $\pm$ 5.42 & 10.06 $\pm$ 2.68 & 8.05 $\pm$ 3.99 & 11.21 $\pm$ 3.68 \\

XGBoost & 13.03 $\pm$ 5.22 & 10.67 $\pm$ 3.97 & 11.44 $\pm$ 3.57 & 11.00 $\pm$ 2.81 & 10.89 $\pm$ 3.32 & 14.36 $\pm$ 4.19 \\

\bottomrule
\end{tabular}
\end{adjustbox}
\label{tab:abl}
\end{center}
\end{table}

We explore the influence of two important factors in our method.
The effect of the different features selected on the results is shown in the Table~\ref{tab:abl}, including NLL only and PPL only. 
Regardless of the meta-model architecture, using one feature alone resulted in larger estimation errors compared to using both features. This corroborates our idea that these two different features reflect the performance of LLM services from different perspectives.

Another ablation study for the number of unlabeled samples is performed on the best meta-model architecture RandomForest. And we chose the best parameters setting obtained by grid search for ablation study with the number of unlabeled samples: the depth of the tree is 10, the number of weak learners is 260, and the sampling ratio is 0.8.

\begin{figure}[t]
\centerline{\includegraphics[width=0.9\linewidth]{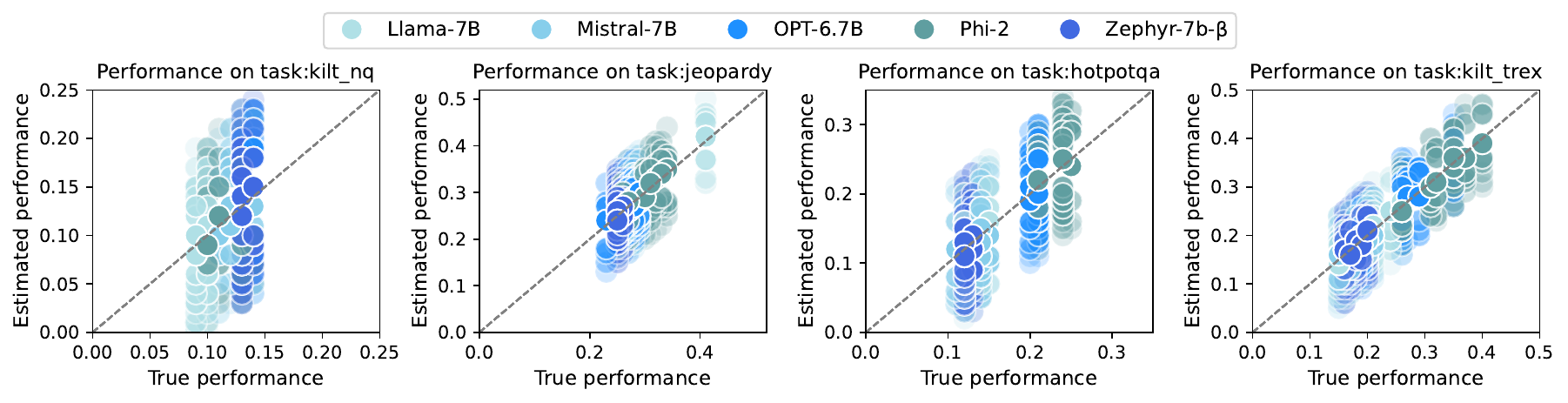}}
\caption{True and estimated performance on four different tasks. The number of unlabeled samples is from 200 to 1600, sampling in the interval of 200. The more opaque the color, the more the unlabeled samples.}
\label{fig:sample}
\end{figure}

For simplicity, we define $n = |T^{*unlabeled}|$ to represent the number of unlabeled samples that need to be used for the estimation of LLM service performance. 
And we performed experiments on all the tasks, reducing the average MAE from 6.30 for $n=200$ to 2.52 for $n=1600$. This effect is visually shown in Figure~\ref{fig:sample}, where increasing $n$ achieves better estimations, despite the need to perform additional LLM service invocations on unlabeled samples. In practice, we recommend 400 unlabeled samples for a task, which can accurately estimate the LLM service performance on the basis of controlling the invocation cost.

\section{Related Work}

\label{sec:rel}
\subsection{Language Model as a Service}

Large language models represent the latest development of generative AI and have shown outstanding performance in the service field, due to their outstanding natural language understanding and representation capability~\cite{11}. 
It is shown to have an attractive capability to "learn"~\cite{BrownMRSKDNSSAA20, chowdhery2022palm}, that is, to perform unlearned tasks correctly given only a few labeled examples. 

However, a rapidly growing number of LLM services have different costs and qualities, resulting in the heterogeneity of the execution performance of the same task~\cite{fragual}. On the one hand, many studies have explored the problem of LLM service selection and composition~\cite{Wang2024ASO,7,51}, to obtain a more affordable and accurate solution to the invocation of LLM services. On the other hand, the performance of LLM services also strongly depends on the choice of prompt templates and examples~\cite{zhao2021calibrate, perez2021true}. The selection~\cite{37} and enhancement of the prompts~\cite{39} have been widely discussed to enhance the performance and generalization of LLM services.

Among all these works, the estimation of the LLM service performance is key because it gives an indication that can be quantified and compared. And this indication can be used to give guidance in the selection of services, order of invocations, optimization of prompts, and so on.

\subsection{LLM Performance Estimation}

LLM performance estimation aims to estimate LLM performance on a specific task in advance. Unlike evaluation~\cite{evaluation}, performance estimation occurs before invocation and focuses on unlearned datasets (out-of-distribution predictions)~\cite{guillory2021predicting}.

Early work focused mainly on estimation based on labeled data, and it is a straightforward idea to take a subset of the dataset and design experiments or benchmarks for performance prediction~\cite{abs230506474, singhal2022assessing}. However, these methods are limited by the representativeness of the data and the cost of annotation. 
In recent years, with the increasing interest in unsupervised learning~\cite{KryeziuS22}, researchers explore how to utilize the LLM internal information for performance estimation~\cite{HuangWLWSWYZ23, li2022estimating}.
These approaches can bypass the need for labeled data, estimating the performance by analyzing LLMs' hidden states or attention weights.
approaches without labeled data are especially effective in domain tasks~\cite{GargBLNS22,10207589}, which learn model confidence to improve the performance estimates for specific tasks. But these approaches require the model to disclose internal details and have limitations for most LLM services that are published in black-box form.

Previous studies provide us with rich experience and enlightenment to explore the LLM service performance estimation methods that do not rely on the labeled dataset. Our approach follows the idea of exploring the available features revealed when scaling on a wide range of unlearned unlabeled tasks, and based on this to estimate the service performance.

\section{Conclusion}

\label{sec:con}
In conclusion, this paper presents a promising approach to addressing the challenge of estimating LLM service performance without labeled data, which can be conveniently applied in a "plug-and-play" manner to a variety of LLM services and tasks. 
By leveraging the meta-model based approach integrated with the ICL paradigm, our method offers accurate performance estimates that exceed baselines, facilitating informed decision-making in LLM service selection and optimization.
Our work still has limitations that need to be explored, such as how to extend the method to larger models and more complex tasks, and how to leverage the ICL capabilities to enhance the generalization of meta-models. We believe that our contributions pave the way for the application of LLM services in practical scenarios, and we look forward to further research based on that.

\subsubsection{\ackname} The work is supported by 
the National Key R\&D Program of China (Grant No.2022YFF0902703), 
the National Natural Science Foundation of China (Grant No.62472121), 
the National Natural Science Foundation of China (Grant No. 62306087), 
and the Special Funding Program of Shandong Taishan Scholars Project.

\newpage
\bibliographystyle{splncs04}
\bibliography{ref}

\end{document}